\documentclass[journal,10pt]{IEEEtran}
\usepackage{amssymb}
\usepackage{times,amsmath,epsfig}
\usepackage{color}
\usepackage{ulem}
 \ifCLASSINFOpdf

\else

\fi

\begin{document}
%

\title{\huge Multi-Feedback Successive Interference Cancellation for Multiuser MIMO Systems}

\author{Peng Li,~\IEEEmembership{Member,~IEEE,}
        Rodrigo C. de Lamare,~\IEEEmembership{Senior Member,~IEEE,}
        and Rui Fa,~\IEEEmembership{Member,~IEEE} 
\thanks{Peng Li and Rodrigo C. de Lamare \{pl534;rcdl500\}@ohm.york.ac.uk are with Department of Electronics, The University of York, Heslington, York, YO10 5DD, UK. Rui Fa R.Fa@liverpool.ac.uk is with the University of Liverpool, UK.}}


%

\maketitle

\begin{abstract}

In this paper, a low-complexity multiple feedback successive interference cancellation (MF-SIC) strategy is proposed for the uplink of multiuser multiple-input multiple-output (MU-MIMO) systems. In the proposed MF-SIC {algorithm with shadow area constraints (SAC)}, an enhanced interference cancellation is achieved by introducing {constellation points as the candidates} to combat the error propagation in decision feedback loops. We also combine the MF-SIC with multi-branch (MB) processing, which achieves a higher detection diversity order. For coded systems, a low-complexity soft-input soft-output (SISO) iterative (turbo) detector is proposed based on the MF and the MB-MF interference suppression techniques. The computational complexity of the MF-SIC is {comparable to} the conventional SIC algorithm {since very little additional complexity is required}. {Simulation} results show that the algorithms significantly outperform the conventional SIC scheme and {approach} the optimal detector.

\end{abstract}


%
\IEEEpeerreviewmaketitle

\section{Introduction}

\IEEEPARstart{I}{n} MU-MIMO systems, the optimum maximum likelihood detection (MLD) scheme which performs an exhaustive search of the constellation map has exponential complexity with the increasing number of data streams (users) \cite{Paulraj}. Therefore, the investigation of sub-optimal low-complexity detection schemes for MU-MIMO systems that can approach the optimal performance is of fundamental importance. The tree search-based sphere decoding (SD) can successfully separate each data stream with reduced complexity compared with the MLD \cite{vikalo}. However, the SD still has an exponential lower bound in complexity for a high number of data streams \cite{Jalden}. {Linear detection (LD) \cite{Paulraj} based on the minimum mean-square error (MMSE) or the zero-forcing (ZF) criteria is a} low-complexity scheme but the error performance is unacceptable due to the multiple access interference (MAI). On the other hand, non-linear detection techniques such as the successive interference cancellation (SIC) used in the vertical Bell Labs layered space-time (V-BLAST) \cite{Foschini1} have a low-complexity, while achieving a reduced MAI than {their} linear counterparts. However, {these decision-driven} detection algorithms suffer from error propagation and performance degradation.

In this paper, inspired by the error propagation mitigation in decision feedback detection \cite{chiani}, \cite{reuter}, \cite{delamare_spa}-\cite{Rui}, {we introduce a novel multiple feedback SIC algorithm with a shadow area constraints (MF-SIC) strategy for detection of multiple users {which requires low computational complexity}. The MF selection algorithm searches several constellation points rather than one constellation point in the conventional SIC algorithm by choosing the most appropriate point in the decision tree. Subsequently, we select this appropriate constellation point as the feedback. By doing so, more points in the decision tree are considered and the error propagation is efficiently mitigated. The selection procedure is constrained to one selected symbol in each spatial layer, unlike sphere decoders which {employ a search procedure} for more layers that increases the computational load. Furthermore, the shadow area constraint (SAC) further saves computational complexity by evaluating the quality of decisions and avoiding unnecessary multiple feedback procedures for reliable {estimates}.}

The MF-SIC is also combined with a multi-branch (MB) \cite{Rui} processing framework to further improve the performance. An iterative receiver (turbo) structure is developed for coded systems and a low-complexity soft-input soft-output (SISO) detector is proposed based on the MF-SIC scheme. Simulation results show that the proposed schemes significantly outperform the conventional SIC schemes and have a comparable performance with the single-user bound.

The contributions of this paper can be summarized as follows: 1) A novel low-complexity {MF-SIC detector} is developed. 2) The MB processing is incorporated into the proposed MF-SIC to achieve a higher detection diversity order and to yield a close to optimal performance. 3) An iterative detection and decoding (IDD) receiver is introduced to approach the MAI free performance in coded systems. 4) A study of the proposed MF-SIC and some existing detection schemes for MU-MIMO systems {is conducted}.

The organization of this paper is as follows. Section II briefly describes the MU-MIMO system model. Section III is devoted to represent the novel MF-SIC scheme as well as its MB processing. Section IV introduces the proposed iterative scheme for coded uplink systems. Section V presents the simulation results and Section VI draws the conclusions of the paper.

\section{System and Data Model}

In this section, the mathematical model of a MU-MIMO system is given. An uplink system with $N_R$ receive antennas at an access point (AP) and $K$ users equipped with a single antenna at the transmitter end is considered. At each time instant, the users transmit $K$ symbols which are organized into a $K \times 1$ vector ${\boldsymbol s}[i] = \big[ s_1[i], ~s_2[i], ~ \ldots, ~s_k[i],~ \ldots,~ s_{K}[i] \big]^T$ and each entry is taken from a modulation constellation $\mathcal{A} = \{ a_1,~a_2,~\ldots,~a_C \}$, where $(\cdot)^T$ denotes transpose and $C$ denotes the number of constellation points. The symbol vector ${\boldsymbol s}[i]$ is then transmitted over flat fading channels and the signals are demodulated and sampled at the receiver. The received signal after demodulation, matched filtering and sampling is collected in an $N_R \times 1$ vector ${\boldsymbol r}[i] =
\big[ r_1[i], ~r_2[i], ~ \ldots,~ r_{N_R}[i] \big]^T$, with sufficient statistics for detection and given by
\begin{equation}
{\boldsymbol r}[i] = \sum_{k=1}^K{\boldsymbol h}_ks_k[i] + {\boldsymbol v}[i] = {\boldsymbol H} {\boldsymbol s}[i] +
{\boldsymbol v}[i],
\end{equation}
where $s_k[i]$ is the transmitted symbol for user $k$, the $N_R \times 1$ vector ${\boldsymbol v}[i]$ is a zero mean complex circular symmetric Gaussian noise with covariance matrix $E\big[ {\boldsymbol v}[i] {\boldsymbol v}^H[i] \big] = \sigma_v^2 {\boldsymbol I}$, where $E[ \cdot]$ stands for expected value, $(\cdot)^H$ denotes the Hermitian operator, $\sigma_v^2$ is the noise variance and ${\boldsymbol I}$ is the identity matrix. The term $\boldsymbol h_k$ represents the $N_R \times 1$ vector of channel coefficients of user $k$, and ${\boldsymbol H}$ is the matrix of the channel vectors for all users. The symbol vector ${\boldsymbol s}[i]$ has zero mean and a covariance matrix $E\big[ {\boldsymbol s}[i] {\boldsymbol s}^H[i]\big] = \sigma_s^2 {\boldsymbol I}$, where $\sigma_s^2$ is the signal power. The model (1) is used repeatedly to transmit a stream of data bits which are separated into blocks representing uses of channels. For a given block, the symbol vector for each user $\boldsymbol{s}_k$ is obtained by mapping it into {the vector $\boldsymbol{x}_k = {[x_{k,1}},..., x_{k,j},..., x_{k,J}]$ with the coded bits}.

\section{Proposed MF-SIC Detector Design}

This section is devoted to the {description of the proposed MF concept and its multi-branch processing framework.} 

%

\subsection{The Multi-Feedback Design}

\begin{figure}[!htb]
\begin{center}
\def\epsfsize#1#2{0.9\columnwidth}
\epsfbox{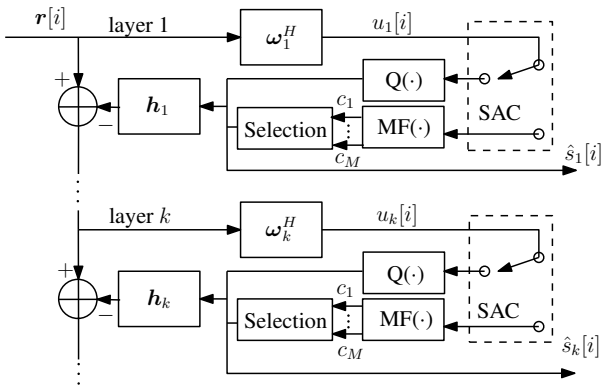} \caption{\footnotesize A reliable interference
cancellation is performed with the MF-SIC scheme. The SAC determines
the reliability of the filter output, the function MF$(\cdot)$
generates $\mathcal{L}=[c_1,\ldots,c_M]$. } \label{MF}
\end{center}
\end{figure}

The structure of the MF-SIC scheme is depicted in Fig.\ref{MF}. The structure considers the feedback diversity by using a number of selected constellation points as the candidates when a previous decision is determined unreliable. In order to find the optimal feedback, a selection algorithm is introduced. This selection scheme prevents the search space from growing exponentially. The reliability of the previous detected symbol is determined by the SAC, which saves the computational complexity by avoiding redundant processing with reliable decisions.

In the following, we only describe the procedure for detecting $\hat{s}_{k}[i]$ for user $k$. The detection of other user streams $\hat{s}_{1}[i]$, $\hat{s}_{2}[i]$,\ldots,$\hat{s}_{K}[i]$ can be obtained accordingly. The soft estimation of the $k$-th user is obtained by $u_{k}[i]=\boldsymbol{\omega}_{k}^H \check{\boldsymbol r}_{k}[i]$ where the $N_r \times 1$ MMSE filter vector is given by $\boldsymbol{\omega}_{k} = (\boldsymbol{{\bar{H}}}_k{{\boldsymbol{{\bar{H}}}_k}}^H+{\sigma_v^2}\boldsymbol{I})^{-1}\boldsymbol{h}_k$ and $\boldsymbol{{\bar{H}}}_k$ denotes the matrix obtained by taking the columns $k,k+1,\ldots K$ of $\boldsymbol {H}$ and $\check{\boldsymbol r}_{k}[i]$ is the received vector after the cancellation of previously detected $k-1$ symbols. For each user, the soft estimation $u_{k}[i]$ is checked by the SAC, which decides whether this decision is reliable according to the metric
\begin{equation}
d_{k} = |u_{k}[i]-a_f|,
\end{equation}
where $a_f$ denotes the constellation point which is the nearest to the soft estimation $u_{k}[i]$ of the $k$-th user, described as
\begin{equation}\label{SAC}
a_f=\arg \min_{a_f\in\mathcal{A}}\{|u_{k}[i]-a_f|\}.
\end{equation}
If $d_{k} > d_{th}$ where $d_{th}$ is the {predefined} threshold, we say it was dropped into the shadow area of the constellation map and this decision was determined unreliable. 
In the presence of the SAC, significant additional computational complexity is saved, the MF-SIC scheme has a comparable complexity to the conventional SIC scheme, as verified by our {studies}.


\subsubsection{Decision Reliable}

If the soft estimation $u_{k}[i]$ is considered reliable, a hard slice will be performed in the same way as in the conventional SIC scheme, the estimated symbol for each data stream $\hat {s}_{k}$ is obtained by $\hat s_k[i] = Q(u_{k}[i])$ {where $Q(\cdot)$ is the signal quantization.} The sliced symbol is considered as a reliable decision for user $k$.

\subsubsection{Decision unreliable}

If the soft estimation is determined unreliable, a candidate vector is generated. $\mathcal{L}=[c_{1}, c_{2},\ldots,c_m,\ldots,c_{M}] \subseteq \mathcal{A}$ is a selection of the $M$ nearest constellation points to the soft estimation $u_{k}[i]$. The size of $\mathcal{L}$ can be either fixed or determined by the signal-to-noise ratio (SNR). A higher SNR corresponds to a smaller $M$ which introduces a trade-off between the complexity and the performance. The unreliable decision {$Q(u_{k}[i])$} is replaced by $\hat {s}_{k}[i] = c_{m_{\tiny \mbox{opt}}}$ where $c_{m_{\tiny \mbox{opt}}}$ is the optimal candidate selected from  $\mathcal{L}$.

The benefits provided by the MF algorithm are based on the assumption that the optimal feedback candidate $c_{m_{\tiny \mbox{opt}}}$ is correctly selected. This {selection} algorithm is described as follows:

{In order to find the optimal feedback, a set of selection vectors $\boldsymbol{b}^1, \ldots, \boldsymbol{b}^m, \ldots, \boldsymbol{b}^M$  is defined, the number of these selection vectors $M$ equals the number of constellation candidates we used for each unreliable decision.
{For the $k$-th layer, a $K \times 1$ vector $\boldsymbol{b}^{m} $ consists of the following elements,} (i) Previously detected symbols $\hat{s}_{1}[i],\ldots,\hat{s}_{k-1}[i]$. (ii) $c_m$ a candidate symbol taken from the constellation for substituting the unreliable decision $Q(u_k[i])$ in the $k$-th layer. (iii) By using (i) and (ii) as the previous decisions, the detection of the following layers ${k+1}, \ldots, q, \ldots, K$-th are performed by the nulling and symbol cancellation which is equivalent to a traditional SIC algorithm. Therefore, we have}
\begin{equation}
\begin{aligned}
&\boldsymbol{b}^{m}[i]=\\
&[\hat{s}_{1}[i],\ldots,\hat{s}_{k-1}[i],c_{m},{b_{k+1}^m[i]}\ldots,b_{q}^m[i],\ldots,b_{K}^m[i]]^T,
\end{aligned}
\end{equation}
where {$b_{q}^m[i]$ is a potential decision that corresponds to the use of $c_m$ in the $k$-th layer,}
\begin{equation}
b_{q}^m[i]=Q(\boldsymbol{\omega}_{q}^H\hat{\boldsymbol{r}}_{q}^m[i]),
\end{equation}
{where $q$ indexes a certain layer between the $({k+1})$-th to the $K$-th.}
\begin{equation}
\hat{\boldsymbol{r}}_{q}^m[i] = \check{\boldsymbol{r}}_{k}[i]-\boldsymbol{h}_k c_{m}-\sum_{p=k+1}^{q-1} \boldsymbol{h}_p {b}_{p}^m[i].
\end{equation}

For each user the same MMSE filter vector $\boldsymbol{\omega}_{k}$ is used for all the candidates, which allows the proposed algorithm to have the computational simplicity of the SIC detection. The proposed algorithm selects the candidates according to
\begin{equation}\label{sele}
m_{\tiny \mbox{opt}}=\arg \min_{1 \leq m \leq M} || \boldsymbol{r}[i]-{\boldsymbol{H}}\boldsymbol{b}^m[i]||^2.
\end{equation}

The $c_{m_{\tiny \mbox{opt}}}$ is chosen to be the optimal feedback symbol for the next layer as well as a more reliable decision for the current user.
The algorithm of the proposed MF-SIC is summarized in TABLE I.

\begin{table}[!t]
\centering
    \caption{The MF-SIC algorithm}     
    \label{tab:MB-MF-SIC}
    \begin{small}
        \begin{tabular}{l}
\hline \\

1:\qquad  $\boldsymbol{\omega}_{k} = ({\boldsymbol{{\bar{H}}}}_k{\boldsymbol{{\bar{H}}}}_k^H+\sigma_v^2 \boldsymbol{I})^{-1}\boldsymbol{h}_k, k=1,\ldots,K$\\
2:\qquad{\bfseries{for}} $k = 1$ \bfseries {to} $K$ \bfseries {do} \qquad \qquad  \rm{{\% For each user}} \\
3:\qquad \qquad ${u}_{k}[i]=\boldsymbol{\omega}_{k}^H\check{\boldsymbol r}_{k}[i]$ \\
4:\qquad \qquad \bfseries {if} $d_{k} > d_{th}$, in shadow area \\
5:\qquad \qquad \qquad $\boldsymbol{\mathcal{L}} = [c_{1}, c_{2},\ldots,c_m,\ldots,c_{M}]^T $\\
6:\qquad \qquad \qquad {\bfseries{for}} $m = 1$ \bfseries {to} $M$ \bfseries {do} \rm{{\% Multiple Feedback}} \\
7:\qquad \qquad \qquad \qquad{\bfseries{for}} $q = k$ \bfseries {to} $K$ \bfseries {do} \\
8:\qquad \qquad \qquad \qquad \qquad$\boldsymbol{\hat{r}}_{k}^m[i] = \check{\boldsymbol{r}}_{k}[i]-\boldsymbol{h}_k c_{m}$\\
\qquad \qquad \qquad \qquad \qquad \qquad \qquad $-\sum_{p=k+1}^{q-1} \boldsymbol{h}_p {b}_{p}^m[i]$\\
9:\qquad \qquad \qquad \qquad \qquad$ b_{q}^m[i]=Q(\boldsymbol{\omega}_{q}^H\boldsymbol{\hat{r}}_{q}^m[i])$ \\
10:\qquad \qquad \qquad \qquad{\bfseries{end for}}\\
11:\qquad \qquad \qquad {\bfseries{end for}}\\
12:\qquad \qquad \qquad $\boldsymbol{b}^{m}[i]=[\hat{s}_{1}[i],\ldots,\hat{s}_{k-1}[i],c_{m},{b_{k+1}^m[i]},\ldots$\\
\qquad \qquad \qquad \qquad \qquad \qquad \qquad $,b_{q}^m[i],\ldots,b_{K}^m[i]]^T$\\
13:\qquad \qquad \qquad $m_{\tiny \mbox{opt}}=\arg \min_{1 \leq m \leq M} || \boldsymbol{r}[i]-\boldsymbol{H}\boldsymbol{b}^m[i]||^2$\\
14:\qquad \qquad \qquad ${\hat s}_{k}[i] = c_{m_{\tiny \mbox{opt}}}$\\
15:\qquad \qquad  \bfseries {else}\\
16:\qquad \qquad ${\hat s}_{k}[i] = Q(u_{k}[i])$\\
17:\qquad \qquad \bfseries {end if}\\
18:\qquad \qquad $\check{\boldsymbol{r}}_{k}[i]=\boldsymbol{r}[i] - \sum_{k=1}^{n-1}{\boldsymbol{h}}_k \hat {s}_{k}[i]$\\
19:\qquad \bfseries {end for}\\\\
\hline
    \end{tabular}
    \end{small}
\end{table}

\subsection{MF-SIC with Multi-Branch Processing}

This section presents the structure of the proposed MF-SIC with multi-branch processing (MB-MF-SIC). The MB-MF-SIC structure is developed based on our previous work \cite{Rui} which contains multiple parallel processing branches of SICs with different ordering patterns.


In the $l$-th branch, the MF-SIC scheme successively computes $\boldsymbol{\hat{s}}_l[i]=[\hat{s}_{l,1}[i],\hat{s}_{l,2}[i],\ldots,\hat{s}_{l,K}[i]]^T$, as detailed in the previous subsection. The term $\boldsymbol{\hat{s}}_l[i]$ represents the $K \times 1$ ordered estimated symbol vector, which is detected according to the IC ordering pattern $\boldsymbol{T}_l,l=1,\ldots,L$ for the $l$-th branch. The IC on the received vector $\check{\boldsymbol{r}}[i]$ is given as follows:
\begin{equation}
\begin{cases}\label{inequ}
    \check{\boldsymbol{r}}_{l,k}[i]=\boldsymbol{r}[i], &  k=1,\\
    \check{\boldsymbol{r}}_{l,k}[i]=\boldsymbol{r}[i] - \sum_{j=1}^{k-1}(\boldsymbol{H}')_j \hat{s}_{l,j}[i], & {k\geq 2} ,\\
\end{cases}
\end{equation}
where the transformed channel matrix $\boldsymbol{H}'$ is obtained by $\boldsymbol{H}' = \boldsymbol{T}_l\boldsymbol{H}.$
The term $(\boldsymbol{H}')_k$ represents the $k$-th column of the ordered channel $\boldsymbol{H}'$ and $\hat{s}_{l,k}$ denotes the estimated symbol for each data stream obtained by the MF-SIC algorithm. At the end of each branch we can transform $\boldsymbol{\hat{s}}_l[i]$ back to the original order $\tilde{\boldsymbol{s}}_l[i]$ by using $\boldsymbol{T}_l$ as $\tilde{\boldsymbol{s}}_l[i]=\boldsymbol{T}_l^T\hat{\boldsymbol{{s}}}_l[i]$. Basically, the MB procedure modifies the original cancellation order in a way that the detector obtains a group of different estimated vectors. At the end of the MB structure, the algorithm selects the branch with the minimum Euclidean distance according to
\begin{equation}
l_{\tiny \mbox{opt}} = \arg \min_{1\leq l \leq L} \boldsymbol{J}(l),
\end{equation}
{for each branch,}
\begin{equation}
\boldsymbol{J}(l) = ||\boldsymbol{r}[i] - \boldsymbol{H}\tilde{\boldsymbol{s}}{_l}[i] ||^2 = ||\boldsymbol{r}[i] - \boldsymbol{H}'\hat{\boldsymbol{s}}_l[i] ||^2.
\end{equation}

In the MB-MF-SIC implementation, the metric $\boldsymbol{J}(l)$ of each MF-SIC branch can be obtained directly from \eqref{sele}. The final detected symbol vector is
\begin{equation}
\boldsymbol{\bar{s}}[i]=\tilde{\boldsymbol{s}}_{l_{\tiny \mbox{opt}}}[i]=\boldsymbol{T}_{l_{\tiny \mbox{opt}}}^T\hat{\boldsymbol{{s}}}_{l_{\tiny \mbox{opt}}}[i].
\end{equation}

This MB scheme can bring a close-to-optimal performance, however, the exhaustive search of $L = K!$ branches is not practical. Therefore, a branch number reduction scheme was developed, namely frequently selected branches (FSB) \cite{Rui}. The FSB algorithm builds a codebook which contains the ordering patterns for the most likely selected branches and the required number of branches to obtain a near-optimal performance is greatly reduced.

\section{Processing with Coded Multiuser MIMO Systems}

In this section, we present the proposed {MF-SIC detector for coded systems which employ} convolutional codes with IDD. We show that a reduced number of turbo iterations can be used with the proposed {schemes} as compared to previously reported turbo multiuser detectors \cite{tad} \cite{Karjalainen}.

The receiver consists of the following two stages: a SISO detector and a set of SISO maximum \textit{a posteriori} (MAP) decoders for the corresponding users. These stages are separated by interleavers and deinterleavers. Specifically, the estimated likelihoods of the convolutionally encoded bits are computed by the detector and these estimates are deinterleaved and serve as input to the MAP decoders. The MAP decoder generates \textit{a posteriori} probabilities (APPs) for each user's encoded bits, and therefore the soft estimate of the transmitted symbol is obtained. The process discussed above is repeated in an iterative manner.

At the output of the SISO detector the \textit{a posteriori} log-likelihood ratio (LLR) of the $j$-th convolutionally encoded bit of the $k$-th user's channel coding block is given by,
\begin{equation}
\Lambda_1[x_{k,j}]=\log \frac{P(x_{k,j}=+1|\boldsymbol{r})}{P(x_{k,j}=-1|\boldsymbol{r})},
\end{equation}
Using Bayes's rule, $\Lambda_1(x_{k,j})$ can be rewritten as
\begin{equation}
\begin{aligned}
\Lambda_1[x_{k,j}]&= \log \frac{P(\boldsymbol{r}|x_{k,j}=+1)}{P(\boldsymbol{r}|x_{k,j}=-1)} +\log \frac{P(x_{k,j}=+1)}{P(x_{k,j}=-1)}\\
& = \lambda_1[x_{k,j}]+\lambda_2^p[x_{k,j}], k = 1,\ldots,K,
\end{aligned}
\end{equation}
where the term $\lambda_2^p[x_{k,j}]=\log \frac{P(x_{k,j}=+1)}{P(x_{k,j}=-1)}$ represents the \textit{a priori} information for the coded bits $x_{k,j}$, which is obtained by the MAP decoder of the $k$-th user in the previous iteration. The superscript $^p$ denotes this value is obtained in the previous iteration. For the first iteration we assume $\lambda_2^p[x_{k,j}]=0$ for all users. The first term $\lambda_1[x_{k,j}]$ denotes the \textit{extrinsic} information which is obtained based on the received signal $\boldsymbol {r}$ and \textit{a priori} information $\lambda_2^p[b_{k,{\tau}}]$ where ${\tau} \neq j$. For the detector, the coded bit \textit{extrinsic} LLR for the $k$-th user is obtained as
\begin{equation}\label{extri}
\lambda_1[x_{k,j}]=\log \frac{\sum_{a_c{\in}\mathcal{A}_{k,j}^+}P(u_k|s_k=a_c)\exp(L_a(a_c))}{\sum_{a_c{\in}\mathcal{A}_{k,j}^-}P(u_k|s_k=a_c)\exp(L_a(a_c))},
\end{equation}
where $\mathcal{A}_{k,j}^+$ and $\mathcal{A}_{k,j}^-$ denotes the subsets of constellation $\mathcal{A}$ where the bit $x_{k,j}$ takes the values 1 and 0, respectively. $L_a(a_c)$ denotes the \textit{a priori} symbol probability for symbol $a_c$. Since 
\begin{equation}
 P(u_k | s_k = a_c) = \frac{1}{\pi\sigma_v^2}\exp({\frac{-||u_k-s_k||^2}{\sigma_v^2}}),
\end{equation}
we rewrite \eqref{extri} as
\begin{equation}
\begin{aligned}
& \lambda_1[x_{k,j}]=  \\
& \log \frac{\sum_{a_c{\in}\mathcal{A}_{k,j}^+}\exp(-||u_k-s_k||^2/\sigma_v^2)\prod_{({\tau}\neq{j})}P(x_{k,{\tau}})}{\sum_{a_c{\in}\mathcal{A}_{k,j}^-}\exp(-||u_k-s_k||^2/\sigma_v^2)\prod_{({\tau}\neq{j})}P(x_{k,{\tau}})},
\end{aligned}
\end{equation}
where $P(x_{k,{\tau}})$ is \textit{a priori} probability of a bit $x_{k,{\tau}}$ and obtained by its \textit{a priori} LLR as \cite{xiaodong}
\begin{equation}
 P(x_{k,{\tau}}) = \frac{1}{2}[1+x_{k,{\tau}}\tanh(\frac{1}{2}\lambda_2^p[x_{k,{\tau}}]){].}
\end{equation}

Then $\lambda_1[x_{k,j}]$ is de-interleaved and fed to the MAP decoder of the $k$-th user as the \textit{a priori} information. The MAP decoder calculates {the} \textit{a posteriori} LLR of each code bit by using the trellis diagram as \cite{delamare_spa}
\begin{equation}
\begin{aligned}
\Lambda_2[x_{k,j}]&= \log \frac{P[x_{k,j}=+1|\lambda_1^p[x_{k,j};\mbox{decoding}]}{P[x_{k,j}=-1|\lambda_1^p[x_{k,j};\mbox{decoding}]}\\
& = \lambda_2[x_{k,j}]+\lambda_1^p[x_{k,j}].
\end{aligned}
\end{equation}
The output of the MAP decoder is obtained by the \textit{a priori} information $\lambda_1^p[x_{k,j}]$ and the \textit{extrinsic} information provided by the decoder. The \textit{a posteriori} LLR of every information bit is also collected by the MAP decoder which is used to make the decision of the message bit after the last iteration. The \textit{extrinsic} information obtained by the $K$ MAP decoders is fed back to the SISO detector as the \textit{a priori} information of all users.
At the first iteration, $\lambda_1$ and $\lambda_2^p$ are statistically independent and as the iterations are performed they become more correlated until the improvement through iterations diminishes.

The structure of the proposed {MF-SIC} with soft {cancellation} (MF-SIC-SC) detector is described in terms of iterations. In the first iteration, the \textit{a priori} information provided by the decoder is zero which heavily degrades the performance of parallel interference cancellation (PIC) based detection. {Therefore, instead of using the PIC based soft {cancellation} (SC/MMSE) \cite{xiaodong} \cite{tad}, in our approach, the proposed {MF-SIC} algorithm is used in the first iteration} to calculate the \textit{extrinsic} information and to feed it to the MAP decoders for all the users. {The soft estimates} $u_k[i]$ is used to calculate the LLRs of their constituent bits. We assume $u_k[i]$ is Gaussian, therefore, the soft output of the SISO detector for the $k$-th user is written as \cite{delamare_spa}
\begin{equation}
u_k[i] = V_ks_k[i]+\epsilon_k[i],
\end{equation}
where $V_k$ is a scalar variable which is equal to the $k$-th users amplitude and $\epsilon_k[i]$ is a Gaussian random variable with variance $\sigma^2_{\epsilon_k}$, since
\begin{equation}\label{stats1}
V_k[i] = E[s_k^*[i]u_k[i]]
\end{equation}
and
\begin{equation}\label{stats2}
\sigma^2_{\epsilon_k} = E[|u_k[i]-V_k[i]s_k[i]|^2].
\end{equation}
The estimates of $\hat{V}_k[i]$ and $\hat{\sigma}^2_{\epsilon_k}$ can be obtained by time averages of the corresponding samples over the transmitted packet.

{After the first iteration, the SC/MMSE performs PIC }by subtracting the soft replica of MAI components from the received vector as
\begin{equation}
\check{\boldsymbol{r}}[i] = \boldsymbol{r}[i]-\boldsymbol{H}{\boldsymbol{z}}[i],
\end{equation}
where {${\boldsymbol{z}}[i] = [u_1[i], \ldots, ~u_{k-1}[i], ~0 ,~u_{k+1}[i], \ldots , ~u_K[i]]$} and a filter is developed to further reduce the residual interference as
\begin{equation}
\boldsymbol{\omega}_k[i]=\arg \min_{\boldsymbol{\omega_k}} E\{|s_k[i] - \boldsymbol{\omega}_k^H\check{\boldsymbol{r}}[i]|^2\},
\end{equation}
where the soft output of the filter is also assumed Gaussian. The first and the second-order statistics of the symbols are also estimated via time averages of (\ref{stats1}) and (\ref{stats2}). {The MF-SIC processing is only applied in the first iteration of the IDD receiver, the proposed MF + selection is introduced in the SIC step to yield refined estimation of symbols. As for MB-MF-SIC-SC, the best MF-SIC-SC branch is selected to provide symbol and bit estimates of the coded information.}

\section{Simulations}

The bit-error-rate (BER) performance of the MF-SIC and MB-MF-SIC is compared with the existing detection algorithms with {uncoded systems and a different number of users.} For channel coded systems, we simulate the IDD schemes with the PIC based SC detector \cite{tad} and compare it with the SIC-SC which uses SIC for the first iteration and the SC is performed for the following iterations \cite{woodward} and MF aided SIC-SC detectors (MF-SIC-SC) {as well as its multiple-branch version (MB-MF-SIC-SC)}. The computational complexity for the proposed detection algorithms is also shown in this section.

Let us consider the proposed algorithms and all their counterparts in the independent and identically-distributed (i.i.d) random flat fading model and the coefficients are taken from complex Gaussian random variables with zero mean and unit variance. As the channel code, we employ a convolutional code with the rate $R=0.5$ and constraint length 3. For each user, 497 message bits are encoded with $g = (7,5)_{\tiny \mbox{oct}}$ and 1000 coded bits are interleaved as a transmitting block, these bits are modulated to 500 QPSK symbols with anti-gray coding. {We also assume that the conventional SIC and MF-SIC are ordered by the same decreased SNR for a fair comparison.} 

\begin{figure}[!htb]
\begin{center}
\def\epsfsize#1#2{0.9\columnwidth}
\epsfbox{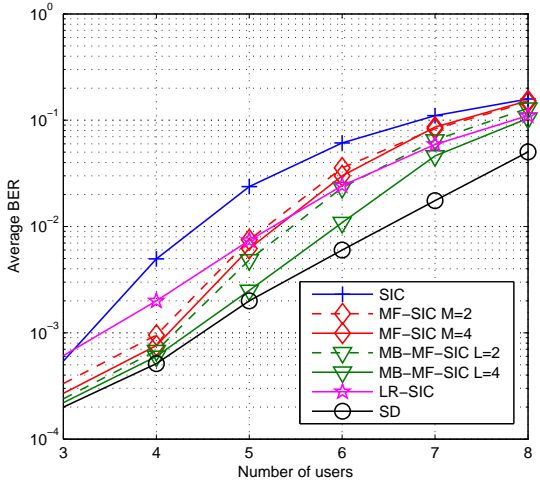} \vspace{-0.3em} \caption{\footnotesize Uncoded
MU-MIMO system with $E_b/N_0 = 12$dB, $N_R = 4$. The proposed MF-SIC
and MB-MF-SIC approach the maximum likelihood performance with 4
users. In the overloaded case, the MB-MF-SIC approaches the SD with
small performance loss with 6 users.} \label{overload}
\end{center}
\end{figure}

{Overloaded systems represent a worst case situation for receivers because of the high level of interference. In practice, it is very unlikely to have a sufficient number of receive antennas for decoupling the spatial signal \cite{kkw}. In Fig.\ref{overload} a system with overloaded transmitting users for uncoded  transmission is considered. }We use $N_R = 4$ receive antennas in the AP and the $E_b/N_0 = 12$ dB at the receiver end.

{The FSB codebook~\cite{Rui} was designed to construct the transformation matrices for MB processing. In this case, [1,2] for $L = 2$ and [1,2,3,5] for $L = 4$. The elements in the FSB codebook indicate the indices of the patterns in the optimum codebook which can be computed by the function \textbf{PERMS()} in MATLAB.} Other parameters are set as $d_{th} = 0.5$, and $M=4$. As for the SD, {we implement the standard SD \cite{MIMOper} to achieve the optimal MLD performance,} the radius $d_{SD}$ is chosen to be a scaled version of the noise variance \cite{vikalo}. The LD and the lattice-reduction aided SIC (LR-SIC) \cite{Cong} are also compared in this plot.

\begin{figure}[!htb]
\begin{center}
\def\epsfsize#1#2{0.9\columnwidth}
\epsfbox{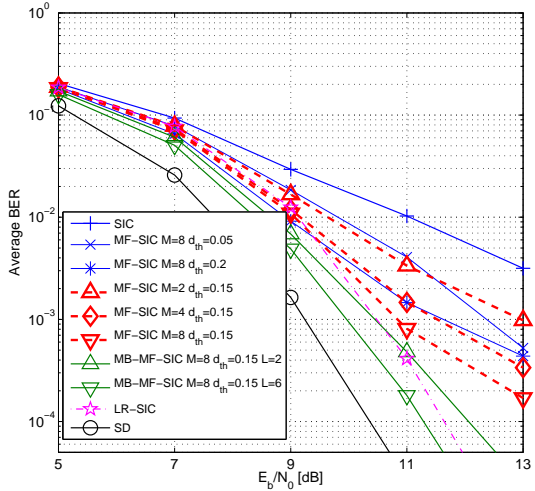} \vspace{-0.3em} \caption{\footnotesize {BER
against SNR for a 4 user system with 16-QAM modulation over flat
fading, the shadow area threshold has an impact on the slop of the
curves.}} \label{16QAM}
\end{center}
\end{figure}

{Another simulation is carried out with a higher level modulation 16-QAM with 4 users. The SNR against BER curves are plotted in Fig.\ref{16QAM}, where we use $M = 2, 4, 8.$ The threshold are $d_{th} = 0.05, 0.15, 0.2$ and we consider a different number of branches $L = 2,6$ for MB-MF-SIC.}

\begin{figure}[!htb]
\begin{center}
\def\epsfsize#1#2{0.9\columnwidth}
\epsfbox{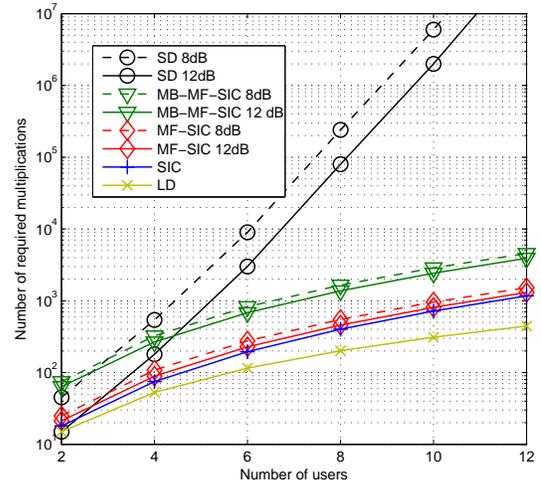} \vspace{-0.3em} \caption{\footnotesize Complexity
in terms of arithmetic operations against the transmit antennas, the
proposed MB-MF-SIC scheme has $L$ times the complexity of the MF-SIC
which has a comparable complexity with the conventional SIC. {$M=4$,
$d_{th}=0.5$, {$L=4$}}.} \label{comp_both}
\end{center}
\end{figure}

In Fig.\ref{comp_both}, the complexity is given by counting the required complex multiplications as the number of users increases. Each MF-SIC branch has a complexity slightly above the SIC while it achieves a significant performance improvement. {We also compared the complexity in terms of the average number of floating-point operations (FLOPS) required per symbol detection by simulations. A simulation performed with the Lightspeed toolbox \cite{lightspeed} and $E_b/N_0 = 12$ dB has shown that for a 16-QAM system with 8 users, the MF-SIC algorithm requires only 2938 FLOPS and a MB-MF-SIC with 9 branches requires 26442 FLOPS while a fixed complexity SD (FSD) \cite{FSD} requires 75120 FLOPS. FSD is one of the lowest complexity SD that we know.}

{For a $4 \times 4$ system and $E_b/N_0 = 16$ dB, the MF-SIC employs the the SAC procedure, the MF concept and the selection algorithm. This leads to the processing of only $6.1\%$ on average over the layers of the estimated symbol with the MF and selection algorithm, whereas for the remaining symbols, the conventional quantization is performed by $\hat s_k[i] = Q(u_{k}[i])$. In terms of processing for each layer, the MF-SIC requires processing $13.34\%$ of the symbols in the first layer, followed by 5.21\%, 2.51\% and 3.15\% for the remaining 3 layers, respectively.}

\begin{figure}[!htb]
\begin{center}
\def\epsfsize#1#2{0.9\columnwidth}
\epsfbox{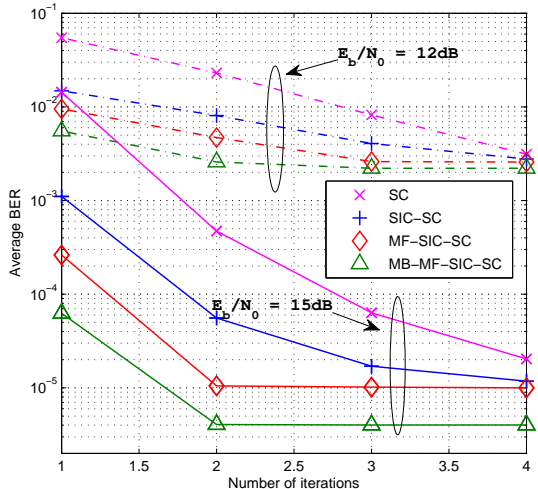} \vspace{-0.3em} \caption{\footnotesize Coded
MU-MIMO system with $N_R=8$ and $K=8$ users. The number of required
iteration is saved in obtaining the converged performance with the
proposed detectors at both $E_b/N_o = 12$ dB and $E_b/N_o = 15$ dB.
{$M=4$, $d_{th}=0.5$, {$L=6$}}.} \label{iter}
\end{center}
\end{figure}

\begin{figure}[!htb]
\begin{center}
\def\epsfsize#1#2{0.9\columnwidth}
\epsfbox{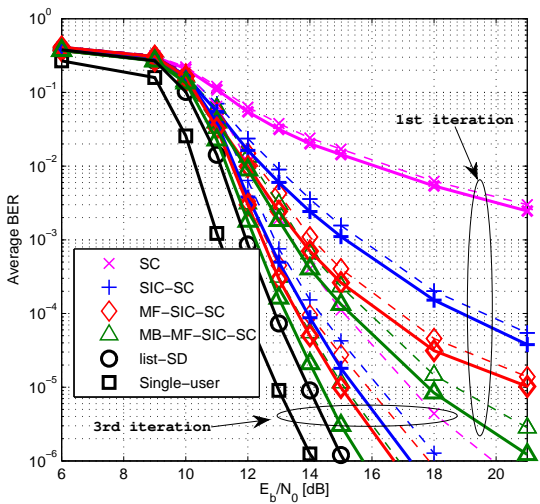} \vspace{-0.3em} \caption{\footnotesize
Convolutional coded system with $K=8$ users. The proposed detectors
have significant performance gains compared with the SC and SIC-SC
detector in their first iteration with both perfect (solid line) and
imperfect channel information (dash line). } \label{IDD}
\end{center}
\end{figure}

For the coded system, the BER against the number of iterations is depicted in Fig.\ref{iter}. We use $K = 8$ times 500 QPSK symbols transmitted over a Rayleigh fading channel, which are collected by $N_R = 8$ antennas. Compared with the previously reported SC and SIC-SC, the proposed MF-SIC-SC and MB-MF-SIC-SC schemes with 2 turbo iterations can obtain a better BER performance than other schemes with 4 iterations, the decoding delay is reduced. Fig.\ref{IDD} shows the simulation with perfect and imperfect channel estimation, where a least {squares} (LS) algorithm is used to estimate the channel weights. We employ a training sequence with 40 symbols which are known at the receiver and the forgetting factor is $\lambda_{LS} = 0.998$. The single-user BER performance describes the performance in an interference free scenario. We can see from this plot that after 3 iterations the slope of the MU-MIMO performance curves are almost the same as the single-user curve with 3 dB (MF $d_{th}=0.5$ $M=4$) and 2 dB (MB-MF $L=6$) performance loss.

\section{Conclusions}

{A low-complexity interference suppression strategy has been developed by introducing multiple constellation points as candidate decisions, and a {cost-effective} selection procedure has been devised to prevent the searching space from growing exponentially.} A multi-branch processing { scheme has also been proposed to enhance the performance of the MF proceeding. Furthermore, we have devised the proposed detectors with IDD and investigate their performance in MU-MIMO systems. The results have shown that proposed iterative detectors approach the single-user performance bound with a lower decoding delay.}


\end{document}